\documentclass{aa}
\usepackage{psfig}
\usepackage{latexsym}

\newcommand{\MC}{\multicolumn}
\newcounter{qub}
\begin{document}
\thesaurus{20(04.19.1; 11.06.2; 11.04.1; 11.19.3; 11.03.2)}
\title{Possibly interacting Vorontsov-Vel'yaminov galaxies }
\subtitle{I. Observations of  VV~432, VV~543 and VV~747 }

\author{Zasov A.V.\inst{1} \and Kniazev A.Y.\inst{2} \and Pustilnik S.A.\inst{2}
\and Pramsky A.G.\inst{2} \and Burenkov A.N.\inst{2} \and Ugryumov A.V.\inst{2}
\and Martin J.-M.\inst{3} }

\offprints{zasov@sai.msu.su}

\institute{State Astronomical Institute of the Moscow State University,
Universitetsky Pr., 13, Moscow, 119899, Russia \\
email: zasov@sai.msu.su
\and
Special Astrophysical Observatory, Nizhnij Arkhyz,
Karachai-Circessia, 357147 Russia \\
email: akn@sao.ru  sap@sao.ru pramsky@sao.ru ban@sao.ru and@sao.ru
\and
D\'epartement de Radioastronomie ARPEGES, Observatoire de Paris,
	      F-92195  Meudon Cedex, France\\
	      email: jmmartin@obspm.fr }

\date{Received \hskip 2cm; Accepted}

\maketitle

\markboth{Zasov et al.: Study of Vorontsov-Vel'yaminov galaxies. I}
{Study of VV galaxies. I.}

\begin{abstract}
Among the galaxies which were included in the Atlas and Catalogue of
Interacting galaxies by Vorontsov-Vel'yaminov (hereafter VV) as multiple
systems (``nests'', ``chains'' and similarly looking systems),
there are many objects, where the interaction is not evident.
Some of them are single objects, including low-mass galaxies with active
star formation  (SF). In this work we present the description of
observations and results of the long-slit spectrophotometry with the
Russian 6\,m telescope of three VV-galaxies looking like double or multiple
systems, and H{\sc i} observations of one of them in order to elucidate
their nature, determine their metallicity, kinematic properties and the
evolution status. Galaxies VV~432 and VV~747 are found to be dwarf systems
with
low oxygen abundance (O/H $\approx$~1/22 and 1/12 of the solar value,
respectively).
Their velocity curves indicate quite  slow rotation with
respective maximum velocities of  about 60 and 80 km~s$^{-1}$,
in agreement with their low luminosities. The distance to VV~432 is
rather uncertain. If it is a member of Virgo Cluster, this is the most
metal-deficient known galaxy of this aggregate.
For galaxy VV~543 the measured emission-line redshift 0.047 appeared ten
times larger than it was given in the original paper and is  cited in
databases. This ``system'' evidently represents an optical pair of two
galaxies with large velocity difference. The fainter western component is an
H{\sc ii}-galaxy, while the brighter one is an absorption-line early-type
galaxy with the radial velocity being 1600 km~s$^{-1}$ lower.

\end{abstract}

\section{Introduction}

Among the definitely interacting galaxies in close pairs or groups,
many of which were discovered by Vorontsov-Vel'yaminov
and included in his Atlas and Catalog of interacting galaxies
(Vorontsov-Vel'yaminov, 1959, 1977) there are  objects  having VV
numbers, in which the
interaction is not so evident.  Such objects were  usually classified by
Vorontsov-Vel'yaminov as the "nests" or often similarly looking systems called
"chains" or "pairs in contact". He considered these objects as compact
fragmenting systems, giving birth to young galaxies, but very soon
observations showed that their nature may be different. Such
systems can present both galaxies of strange and unusual shapes (either
single objects or mergers), and really multiple systems, where it is hard to
guess the number of components  without detailed studies.
In many cases only spectral measurements of  their gas velocity
distribution enable to prove or to disprove their solitude.

Earlier spectral observations confirmed that some VV systems  actually
present single dwarf galaxies with clumpy inner structure, resembling dwarf
irregulars with multiple regions of active star formation (SF), or
blue compact galaxies (BCGs)
(Vorontsov-Vel'yaminov 1979a,b; Afanasiev et al.\, 1980; Arkhipova et al.\,
1981, 1987a,b,c).
Only a few single nearby VV galaxies have been studied in detail
(e.g. VV~556~$\equiv$~GR~8, VV~499~$\equiv$~DDO~053).

The important questions on the  chemical abundances in the studied VV galaxies
left outside the scope of these early observations. New opportunities
appeared due to new CCD-detectors, which enable to realize high
sensitivity and large dynamical range. It induced authors to return back
to the study of "nest-like" VV galaxies. Besides, many new
observational data have appeared on many of these objects.

For the current study we selected more than twenty VV galaxies looking like
multiple systems or singular irregular  systems in the  POSS/DSS
\footnote
{DSS is Digital Sky Survey distributed by Hubble Space Telescope Science
Institute} images,
for which such indicators of interaction as well defined tidal tails or
bridges are absent. The main objectives of our investigation are:

-- to clarify the distances, total luminosities and masses of those
systems, systemic velocities of which are not known or badly known;

-- to carry out high S/N ratio spectrophotometry in order to address
problems of chemical abundances and evolutional status of these objects,

-- to analyse the inner gas kinematics and structural properties of
the objects.

 Since many of VV galaxies in question are
dwarfs with recent or current SF burst, it is also important to
check possible companions which could exert strong enough tidal action.


\begin{table*}
\begin{center}
\caption{\label{Tab1} Journal of observations}
\begin{tabular}{lllcrccc} \\ \hline
\MC{1}{c}{ Galaxy } &
\MC{1}{c}{ Date } &
\MC{1}{c}{ Instrument } &
\MC{1}{c}{ Grating } &
\MC{1}{c}{ Exposure } &
\MC{1}{c}{ Wavelength } &
\MC{1}{c}{ Dispersion } &
\MC{1}{c}{ PA } \\

\MC{1}{c}{ Name } &  &  &
\MC{1}{c}{ [grooves/mm] } &
\MC{1}{c}{ time [s] } &
\MC{1}{c}{ Range [\AA] } &
\MC{1}{c}{ [\AA/pixel] } &
\MC{1}{c}{ [Degree] } \\

\MC{1}{c}{ (1) } &
\MC{1}{c}{ (2) } &
\MC{1}{c}{ (3) } &
\MC{1}{c}{ (4) } &
\MC{1}{c}{ (5) } &
\MC{1}{c}{ (6) } &
\MC{1}{c}{ (7) } \\
\hline
\\[-0.3cm]
VV~432&  12.02.1999  & LSS+PMCCD & 325 &  900 & $3700-8000$ & 4.6 & ~~27 \\
VV~432&  23.04.1999  & LSS+PMCCD & 1302& 2$\times$1200 & $4000-5200$ & 1.2 & ~~27 \\
VV~432&  23.04.1999  & LSS+PMCCD & 1302& 2$\times$1200 & $6000-7200$ & 1.2 & ~~27 \\
VV~543&  11.02.1999  & LSS+PMCCD & 325 &  900 & $3700-8000$ & 4.6 & 101 \\
VV~747&  12.02.1999  & LSS+PMCCD & 325 & 1200 & $3700-8000$ & 4.6 & ~~56 \\
\hline \\[-0.2cm]

\end{tabular}
\end{center}
\end{table*}


\begin{table*}[hbtp]
\centering{
\caption{\label{Tab2} Main parameters of studied VV-galaxies}

\begin{tabular}{lrrrr} \hline
\rule{0pt}{10pt}
Parameter                     & VV~432               & VV~543~W              & VV~543~E            & VV~747    \\ \hline
$\alpha_{2000}$               & 12$^h$17$^m$34.7$^s$ & 13$^h$42$^m$22.2$^s$  & 13$^h$42$^m$23.5$^s$& 10$^h$57$^m$46.9$^s$   \\
$\delta_{2000}$               & +12$^\circ$23$^{\prime}$46$^{\prime\prime}$  & +29$^\circ$49$^{\prime}$33$^{\prime\prime}$ & +29$^\circ$49$^{\prime}$30$^{\prime\prime}$ & +36$^\circ$15$^{\prime}$38$^{\prime\prime}$ \\
B$_{tot}^L$                   & 14.73$\pm$0.15       & 17.7$\pm$0.3          & 15.20$\pm$0.20      & 15.52$\pm$0.71   \\
A$_B^N$                       &  0.03                & 0.00                  & 0.00                & 0.04             \\
V$_{HeI}$ (km/s)              & $-160\pm$6           & 14100$\pm$20          & 12480$\pm$90        & 629$\pm$7$^5$    \\
Dist$_{Vir}$(Mpc)             & 20.7$\pm$0.8         & 188.4$\pm$0.3         & 166.8$\pm$1.2       & 9.2$\pm$0.1      \\
M$_{B}$$^{1}$                 &  $-16.8$             & $-18.6$               & $-20.9$             & $-14.4$          \\
D$_{25}$  (arcsec)            &  111                 & $\approx$16           &  39                 &  50              \\
D$_{25}$  (kpc)               & 11.1                 & $\approx$14.6         &  31.5               &  2.2             \\
Axis ratio b/a$^L$            &  0.31                & 0.50                  & 1.00                & 0.85             \\
12+log(O/H)  \                & 7.58$\pm$0.06        & 8.5$\pm$0.1           & ---                 & 7.85$\pm$0.05    \\
H{\sc i} flux$^{2}$           & 9.04$\pm$1.99$^3$    & ---                   & ---                 & 4.03$\pm$0.39    \\
W$_{20}$~km s$^{-1}$          & 123$^3$              & ---                   & ---                 & 117$\pm$11       \\
M(H{\sc i}) (10$^{8}$M$_{\odot}$) & 9.1$\pm$2.0      & ---                   & ---                 & 0.81$\pm$0.08    \\
M(H{\sc i})/L$_{B}$$^{4}$     & 1.2                  & ---                   & ---                 & 1.0              \\
\hline
\multicolumn{4}{l}{B$_{tot}$ --- total blue magnitude; M$_{B}$ --- absolute blue magnitude} \\
\multicolumn{4}{l}{D$_{25}$ --- Diameter at surface brightness $\mu_{B}$ = 25~$mag/\Box^{\prime\prime}$} \\
\multicolumn{4}{l}{A$_{B}$ --- Galactic extinction; $^{L}$\,\ Data from LEDA; $^{N}$\,\ Data from NED} \\
\multicolumn{4}{l}{$^{1}$\,\ With Galactic extinction correction; $^{2}$\,\ Units of (Jy$\cdot$km/s)} \\
\multicolumn{4}{l}{$^{3}$\,\ from Schneider et al. (\cite{Schneider91}); $^{4}$\,\ In (M/L$_{B}$)$_{\odot}$; $^{5}$\,\ From H{\sc i}-profile} \\
\end{tabular}
}
\end{table*}

In this paper, the first in the series, we present the results of recent
long-slit spectroscopy for three VV objects:
VV~432, VV~543 and VV~747 and the observation in HI-line of VV~747.

In section~\ref{observations} we describe observations, data reduction,
abundances determination and the measurements of velocity distribution
of ionized gas along the slit.

Observations of  VV~747 in the HI-line 21 cm and their results are
presented in section~\ref{HI_observations}.
In section~\ref{Individual_prop} we consider the individual properties
of studied galaxies.
Discussions and preliminary conclusions are presented in section
~\ref{Conclusions}.  We adopt throughout the paper
H$_0 =$75 km sec$^{-1}$.

\section{Spectral observations, data reduction and analysis}
\label{observations}

\subsection{Observations}

The spectroscopic data were obtained  with the
6\,m telescope of  the Special Astrophysical Observatory of Russian
Academy of Science (SAO RAS) during two runs in February and April 1999.
The Long-Slit spectrograph (LSS in Table~\ref{Tab1}) (Afanasiev et al.
\cite{Afanasiev95}) at the telescope prime focus was equipped with a
Photometrics CCD-detector PM1024 (with $24\times24\mu$m pixel size) (PMCCD
in Table~\ref{Tab1}) installed at Schmidt-Cassegrain camera F/1.5.
Most of the long-slit spectra ($1.2\arcsec\times180\arcsec$) were obtained
with the grating of 325 grooves/mm, giving a dispersion of 4.6~\AA/pixel.
Additional data were obtained with
the grating  of 1302 grooves/mm and dispersion 1.2~\AA/pixel.
For the latter set-up the slit of $2\arcsec\times180\arcsec$ was used.
The scale along the slit was 0.39$\arcsec$/pixel.

The resulting resolution  (FWHM) was about $14-15$~\AA\, for the
first set-up, and about 3.7~\AA\, for the second set-up.
Reference spectra of an Ar--Ne--He lamp were recorded before or after
each observation to provide  wavelength calibration.
Spectrophotometric standard stars from Massey et al. (\cite{Massey88})
were observed for flux calibration at least twice a night.

Observations and data processing in this set-up have been
conducted mainly under the software package {\tt NICE} in MIDAS, described by
Kniazev \& Shergin (\cite{Kniazev95}).


\begin{table*}[hbtp]
\centering{
\caption{\label{Tab3} Line intensities in the knot ``a'' of VV~432, VV~747 and VV~543W}
{\small
\begin{tabular}{lcccccccc} \hline
\rule{0pt}{10pt}
   &          \MC{2}{c}{VV~432 (a) (4.6~\AA/$pixel)$ }     &  \MC{2}{c}{VV~432 (a) (1.2~\AA/$pixel)$}                                                                   &\MC{2}{c}{VV~747 (a) (4.6~\AA/$pixel)$ } & \MC {2}{c}{VV~543W (4.6~\AA/$pixel)$ } \\ \hline
\rule{0pt}{10pt}
$\lambda_{0}$(\AA)                      & F($\lambda$)/F(H$\beta$)&I($\lambda$)/I(H$\beta$)     &F($\lambda$)/F(H$\beta$)&I($\lambda$)/I(H$\beta$)                      &F($\lambda$)/F(H$\beta$)&I($\lambda$)/I(H$\beta$)     &F($\lambda$)/F(H$\beta$)&I($\lambda$)/I(H$\beta$)        \\ \hline
3727\  [O\ {\sc ii}]\                   & 2.17$\pm$0.10     & 2.63$\pm$0.12 &     ---        &     ---                                                              & 1.08$\pm$0.06 & 1.57$\pm$0.10     & 3.79$\pm$0.33 & 4.31$\pm$0.40   \\
3868\  [Ne\ {\sc iii}]\                 & 0.33$\pm$0.03     & 0.39$\pm$0.04 &     ---        &     ---                                                              & 0.30$\pm$0.01 & 0.40$\pm$0.02     &     ---       &   ---           \\
4101\  H$\delta$\                       & 0.26$\pm$0.02     & 0.30$\pm$0.04 & 0.23$\pm$0.03  & 0.26$\pm$0.04                                                        & 0.22$\pm$0.01 & 0.28$\pm$0.02     &     ---       &   ---           \\
4340\  H$\gamma$\                       & 0.49$\pm$0.02     & 0.54$\pm$0.03 & 0.50$\pm$0.03  & 0.54$\pm$0.04                                                        & 0.40$\pm$0.01 & 0.47$\pm$0.02     & 0.50$\pm$0.07 & 0.53$\pm$0.10   \\
4363\  [O\ {\sc iii}]\                  & 0.11$\pm$0.01     & 0.12$\pm$0.02 & 0.09$\pm$0.03  & 0.09$\pm$0.03                                                        & 0.09$\pm$0.01 & 0.11$\pm$0.01     &     ---       &   ---           \\
4861\  H$\beta$\                        & 1.00$\pm$0.02     & 1.00$\pm$0.03 & 1.00$\pm$0.04  & 1.00$\pm$0.04                                                        & 1.00$\pm$0.01 & 1.00$\pm$0.02     & 1.00$\pm$0.09 & 1.00$\pm$0.11   \\
4959\  [O\ {\sc iii}]\                  & 1.10$\pm$0.02     & 1.08$\pm$0.02 & 1.09$\pm$0.04  & 1.07$\pm$0.04                                                        & 1.79$\pm$0.18 & 1.74$\pm$0.18     & 0.46$\pm$0.08 & 0.45$\pm$0.08   \\
5007\  [O\ {\sc iii}]\                  & 3.36$\pm$0.05     & 3.28$\pm$0.05 & 3.25$\pm$0.10  & 3.18$\pm$0.09                                                        & 5.49$\pm$0.23 & 5.27$\pm$0.22     & 1.26$\pm$0.11 & 1.23$\pm$0.11   \\
5876\  He\ {\sc i}\                     & 0.09$\pm$0.02     & 0.08$\pm$0.02 &     ---        &     ---                                                              & 0.09$\pm$0.01 & 0.07$\pm$0.01     &     ---       &   ---           \\
6548\  [N\ {\sc ii}]\                   & 0.05$\pm$0.01     & 0.05$\pm$0.01 & 0.03$\pm$0.01  & 0.03$\pm$0.01                                                        & 0.03$\pm$0.01 & 0.02$\pm$0.00     & 0.20$\pm$0.03 & 0.17$\pm$0.02   \\
6563\  H$\alpha$\                       & 3.39$\pm$0.07$^1$ & 2.74$\pm$0.06 & 3.39$\pm$0.10  & 2.77$\pm$0.09                                                        & 2.34$\pm$0.04 & 2.79$\pm$0.05$^2$ & 3.35$\pm$0.23 & 2.88$\pm$0.22   \\
6584\  [N\ {\sc ii}]\                   & 0.14$\pm$0.04     & 0.11$\pm$0.03 & 0.11$\pm$0.01  & 0.09$\pm$0.01                                                        & 0.09$\pm$0.03 & 0.06$\pm$0.02     & 0.74$\pm$0.08 & 0.64$\pm$0.07   \\
6717\  [S\ {\sc ii}]\                   & 0.26$\pm$0.02     & 0.21$\pm$0.01 & 0.29$\pm$0.02  & 0.23$\pm$0.01                                                        & 0.13$\pm$0.01 & 0.09$\pm$0.01     & 0.99$\pm$0.11 & 0.84$\pm$0.10   \\
6731\  [S\ {\sc ii}]\                   & 0.17$\pm$0.01     & 0.13$\pm$0.01 & 0.20$\pm$0.02  & 0.16$\pm$0.01                                                        & 0.10$\pm$0.01 & 0.07$\pm$0.01     & 0.60$\pm$0.10 & 0.51$\pm$0.09   \\
  & &                                                                                                                                                                 &         & \\
C(H$\beta$)\ dex          & \MC {2}{c}{0.275$\pm$0.026} & \MC {2}{c}{0.265$\pm$0.037}                                                                                   & \MC {2}{c}{0.445$\pm$0.016} & \MC {2}{c}{0.190$\pm$0.088} \\
EW(abs)\ \AA\             & \MC {2}{c}{0.100$\pm$1.807} & \MC {2}{c}{0.050$\pm$2.067}                                                                                   & \MC {2}{c}{0.080$\pm$0.985} & \MC {2}{c}{0.100$\pm$0.827} \\
F(H$\beta$)               & \MC {2}{c}{123.10$\pm$1.81} & \MC {2}{c}{91.08$\pm$2.54}                                                                                    & \MC {2}{c}{175.09$\pm$1.60} & \MC {2}{c}{7.78$\pm$0.50}  \\
EW(H$\beta$)\ \AA\        & \MC {2}{c}{120.96$\pm$1.78} & \MC {2}{c}{128.37$\pm$3.58}                                                                                   & \MC {2}{c}{125.46$\pm$1.14} & \MC {2}{c}{13.52$\pm$0.87}  \\

\hline

\multicolumn{9}{l}{\rule{0pt}{10pt}$^1$ Due to calibration problems the intensity of H$\alpha$ is likely a bit overestimated. To check its effect we redone calculation of O/H} \\
\multicolumn{9}{l}{for C(H$\beta$)=0. It increases 12+lg(O/H) by 0.01. $^2$ --- the intensity of H$\alpha$ is recalculated from the recombination ratio to H$\beta$.} \\
\end{tabular}
} }
\end{table*}


\subsection{Data reduction and abundances determination}
\label{Abund_determ}

The data  reduction was performed in SAO RAS, using various packages of MIDAS
(see Kniazev et al. (\cite{Kniazev2000}) for details).

In Table~\ref{Tab2} we summarize the main observational parameters of the
discussed three VV ``nests''.  They include the names of the objects, their
coordinates for the epoch J2000, the apparent blue magnitudes and the
corresponding references, the radial heliocentric velocities, measured in
this work with their r.m.s. uncertainties, maximal angular sizes,
absolute blue magnitudes and the oxygen abundances (12+log(O/H)).

\begin{figure*}[hbtp]
\vspace*{-1.0cm}
\psfig{figure=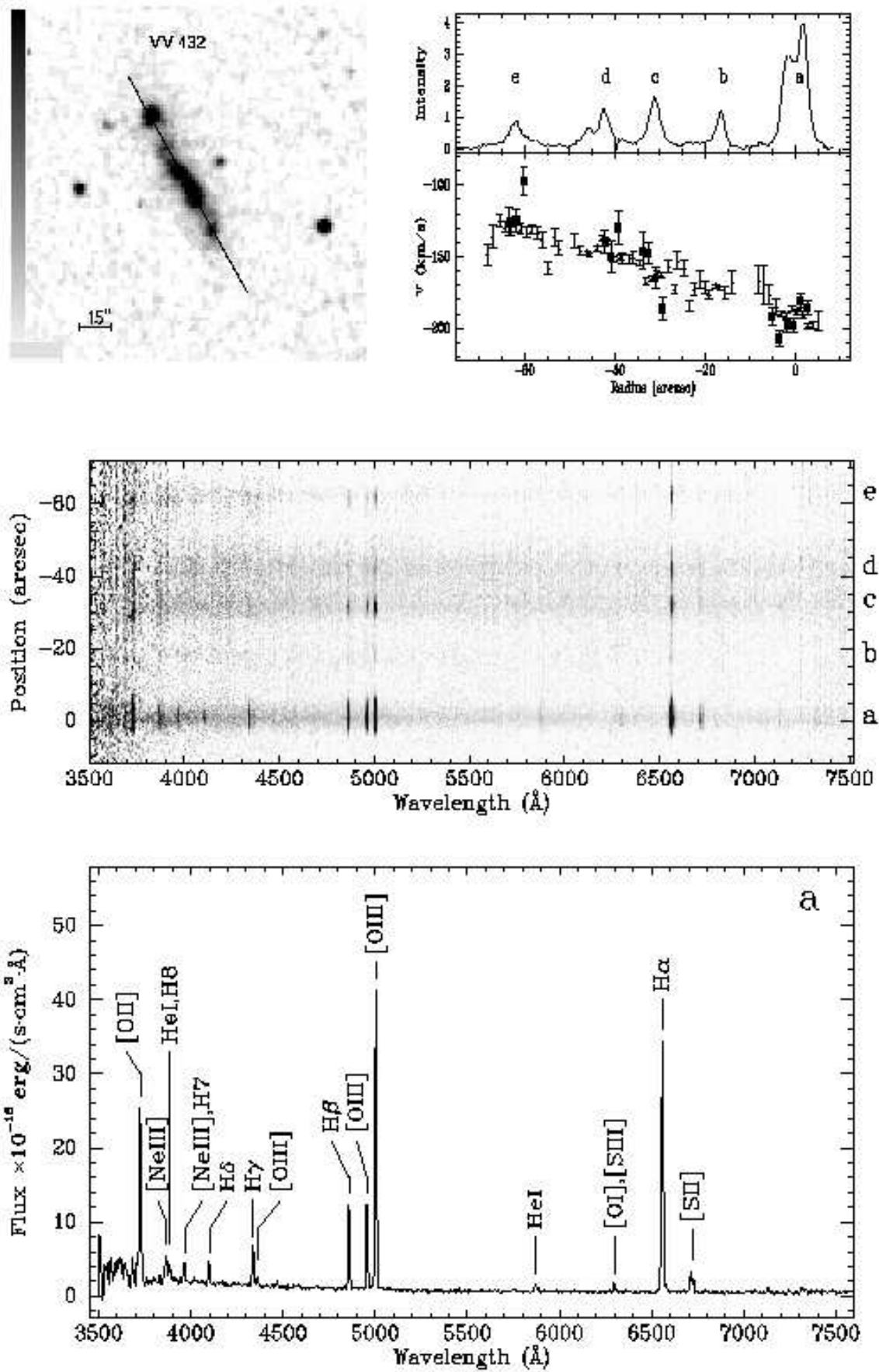,angle=0,width=18cm}
\vspace*{-1cm}
\caption[c]{From top-left to bottom:
{\bf a)} DSS image of galaxy VV~432 with the position of
long slit superimposed. The spot ``a'' is NE bright region at the
edge of the galaxy;
{\bf b)} Brightness profile of H$\alpha$ in relative intensities
along the slit and the velocity curve. Only independent along the slit
points (accounting for seeing) from the spectra
with dispersion 1.2~\AA/pixel and 4.6~\AA/pixel are shown by
arrows and filled boxes respectively. The error bars correspond to
$\pm 1\sigma$.
{\bf c)} 2-D spectrum of VV~432 with dispersion 4.6~\AA/pixel;
{\bf d)} 1-D spectrum of the brightest NE spot (``a'')}
\label{VV432_fig}
\end{figure*}

\begin{figure*}[hbtp]
\vspace*{-1.0cm}
\psfig{figure=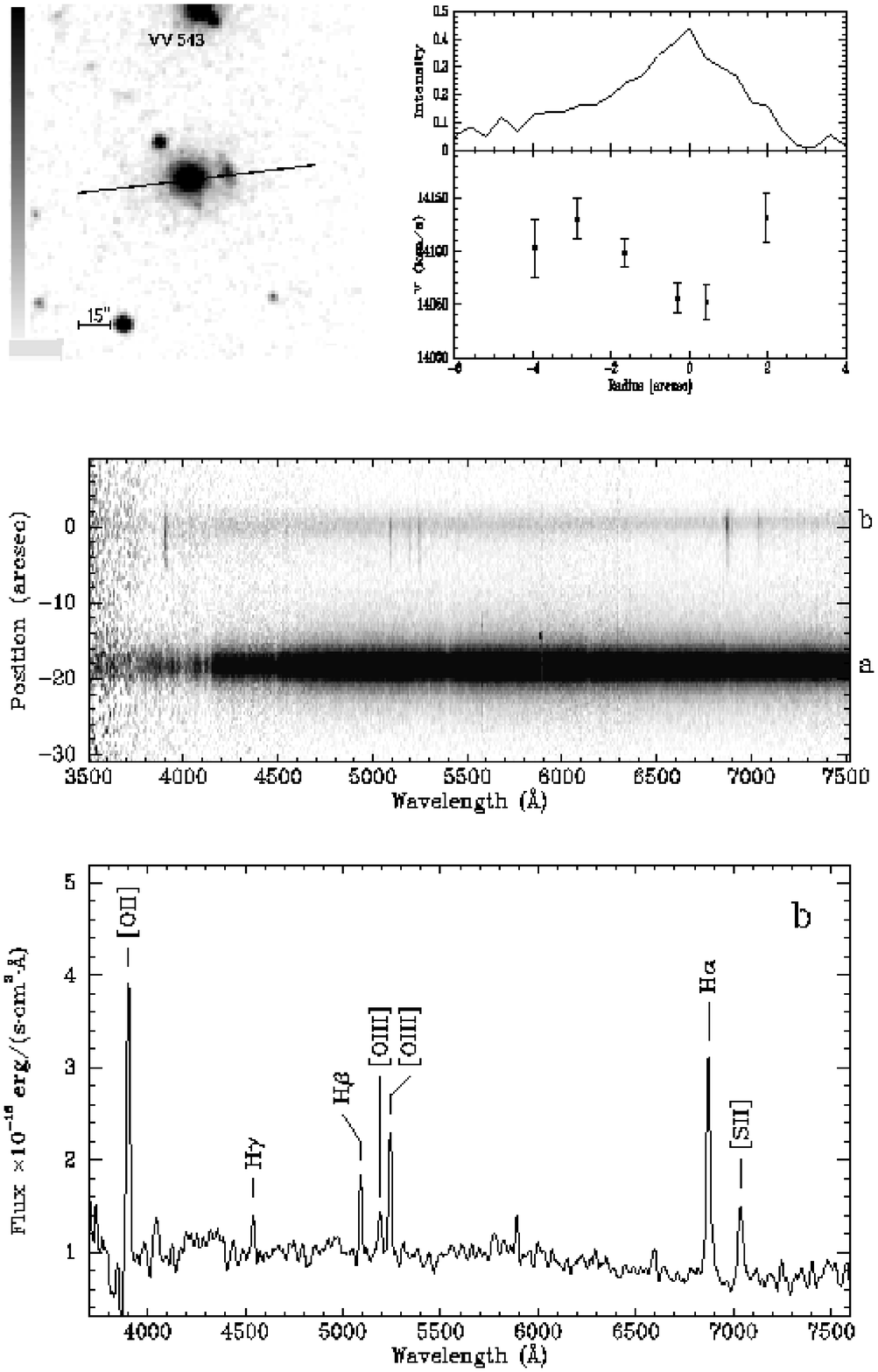,angle=0,width=18cm}
\caption[c]{From top-left to bottom:
{\bf a)} DSS-1 image of galaxy pair VV~543E and VV~543W with the
position of long slit superimposed;
{\bf b)} Brightness profile of H$\alpha$ in relative intensities
along the slit and the velocity curve.;
{\bf c)} 2-D spectrum of the pair VV~543;
{\bf d)} 1-D spectrum of the Western component
}
\label{VV543_fig}
\end{figure*}

Direct images of studied galaxies, extracted from the DSS and
the position of long slit, indicated by bar are presented in Fig.~\ref{VV432_fig}a,
\ref{VV543_fig}a, \ref{VV747_fig}a. Corresponding
2-D spectra are shown in Fig.~\ref{VV432_fig}c, \ref{VV543_fig}c,
\ref{VV747_fig}c.  The brightness profiles of
H$\alpha$ line along the slit, and corresponding velocity curves are
illustrated
in Fig.~\ref{VV432_fig}b, \ref{VV543_fig}b, \ref{VV747_fig}b.
In Fig.~\ref{VV432_fig}d, \ref{VV543_fig}d,  and
\ref{VV747_fig}d we present 1-D spectra,
extracted from 2-D spectra, which were used for the measurements of line
intensities, determination of physical
conditions  and abundances of H{\sc ii}-regions.

The resulting observed emission line intensities
$F(\lambda)$ of various ions relative to H$\beta$,
both uncorrected and corrected for interstellar extinction and underlying
stellar absorption $I(\lambda)$ (following the procedure described
by Izotov et al.~\cite{Izotov97}) for the brightest parts of the galaxies
are presented in Tables~\ref{Tab3}
along with the extinction coefficient C(H$\beta$),
the equivalent width of absorption Balmer hydrogen lines EW(abs),
the equivalent width of H$\beta$ line EW(H$\beta$)
and the observed H$\beta$ flux.

For the abundances determination we used
the scheme, described in detail by Izotov et al. (\cite{Izotov94},
\cite{Izotov97}).
The electron temperatures and densities in H{\sc ii}-regions of the observed
VV-galaxies and their abundances  of O, N and Ne are
summarized in Table~\ref{Tab6}.

For VV~543 the [O{\sc iii}]-line 4363~\AA\ is not detected. Therefore to
estimate  its metallicity we employ the empirical method (see
e.g. Pagel et al. (\cite{Pagel79}), McGaugh (\cite{McGaugh91}) and Olofsson
(\cite{Olofsson97})).  Its uncertainty for 12+log(O/H) can be
as large as 0.2--0.3 dex. We also applied this empirical method to all
four   individual  knots of VV~432.

The ratio [O{\sc iii}]/[N{\sc ii}] of extinction corrected intensities
of the lines $\lambda$\,4959,5007~\AA\ and $\lambda$\,6548,6584~\AA\
enable one to avoid well known ambiguity of the empirical method
(Alloin et al. \cite{Alloin79}).
Taking into account these ratios,
one can use $R_{23}$ and $Q$ parameters to get more
reliable estimate of $O/H$ through the model curves by Olofsson
(\cite{Olofsson97}). By this way it was obtained for VV~543W the
abundance 12+log(O/H) =
8.5$\pm$0.1, which is about 0.3 lower than the value, derived
from McGaugh (\cite{McGaugh91}) tracks, and seems to be in a better
agreement with the abundance, expected for underluminous
H{\sc ii}-galaxies.


\subsection{Velocity curves}

To obtain line-of-sight velocity distribution along the slit  we
use the MIDAS programs, kindly presented to authors by D. Makarov.

To increase the accuracy of
derived rotation curves the programs were modified to include additional
corrections using close lines in the reference spectrum.

The procedure includes the following steps:
1) a linearisation of 2-D spectrum of the object and of the reference spectrum;
2) a measurement of the position of the H$\alpha$ emission for each of the 250
position rows using gaussian fitting;
3) an estimation of the background and  S/N ratio for the
H$\alpha$ emission in each row;
4) a similar measurement of the nearby line Ne{\sc i}~$\lambda$\,6598.95~\AA\
in the reference spectrum and compiling the table of
differences between the laboratory wavelength of this line and the
measured one for each row;
5) a polynomial fitting of these differences and a determination of
the residual scattering (r.m.s.);
6) the application of this fitting polynomial to correct
the measured wavelengths of H$\alpha$. The resulting error of
wavelengths measurements is determined by quadratic summing of the
fitting r.m.s., found in the previous step, and the measurement error
of velocity for each point
along the slit, estimated from gaussian fitting. The latter varies
between 2.5 km~s$^{-1}$ for the points with the highest signal and 9
km~s$^{-1}$ for the points with low S/N ratio for dispersion 1.2~\AA/pixel;
7) rebinning of H$\alpha$ data along the slit,
corresponding to the seeing during the observations (4 or 5 pixels in April
and February, respectively).

Below we use only those velocity estimates which
satisfy the criteria S/N~$>$~3 and $\sigma_V$~$<$ 15~km~s$^{-1}$.
Resulting high accuracy of corrected observed wavelength enables one
to study irregularities of the velocity curve with an amplitude as low as
10 km~s$^{-1}$ on the angular scale up to 20\arcsec.

For VV~432, the rotation curves which were obtained with the low and
high resolution spectra are presented in
Fig.~\ref{VV432_fig}b. Their similarity shows that our low dispersion
spectra can be used to derive preliminary dynamical parameters
of the studied galaxies.

\section{H{\sc i} data for VV~747}
\label{HI_observations}

For VV~747, the interpretation  of the optical data remains
ambiguous regarding its multiplicity.
(see Sect.~\ref{747_mult}).
In order to settle this ambiguity we investigated the dynamical
state of the system and performed single-dish H{\sc i} observations.

\subsection{Observations and data reduction}

The 21~cm H{\sc i}-line observations were carried out in July 1999 with the
Nan\c {c}ay\footnote{The Nan\c {c}ay Radioastronomy Station is part of the
Paris Observatory and is operated by the Minist\`ere de l'Education
Nationale and Institut des Sciences de l'Univers of the Centre National
de la Recherche Scientifique.} 300\,m telescope (NRT).
The NRT has a half-power beam width of
3.7$^{\prime}$ (EW)~$\times$~22$^{\prime}$ (NS) at the declination
$\delta = 0^\circ$.

A dual polarization receiver has been used, with a system temperature 
of $\approx$~40~K in the horizontal and vertical linear 
polarizations. The gain of the telescope was 1.1 K Jy$^{-1}$ at a 
declination of $\delta = 0^\circ$. The observations were made in
position switching mode with 1-minute on-source and 1-minute off-source 
integrations.

Since VV~747 had a known optical redshift, we split the 
autocorrelator into two, each bank covering a bandwidth of 6.4 MHz, 
and centred at the frequency corresponding to the optical redshift. 
The velocity coverage was 1\,350 km~s$^{-1}$.  The channel
width was 2.6 km~s$^{-1}$ and  after averaging pairs of adjacent
channels the effective resolution was 6.3 km~s$^{-1}$.

The data has been reduced using the software developed by the 
telescope's staff. Both polarization spectra were calibrated and 
processed independently, and were averaged to improve the 
signal-to-noise ratio. Errors were calculated following
Schneider et al. (\cite{Schneider86}).

\subsection{Results. H{\sc i}-profile and its parameters}
\label{HI_res}

\begin{figure}[hbtp]
\hspace*{-0.5cm}
\psfig{figure=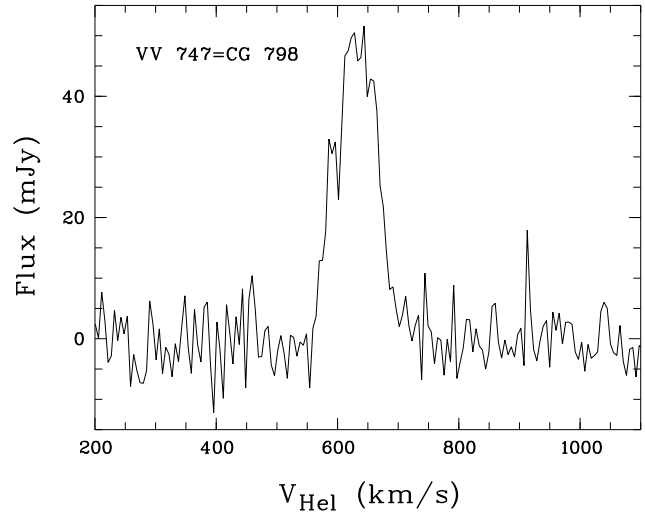,angle=-90,width=10.5cm}
\caption[c]{H{\sc i}-profile of galaxy VV~747. Velocity resolution
	    is 6.3~km~s$^{-1}$, W$_{0.5}$=90~km~s$^{-1}$,
	    W$_{0.2}$=117~km~s$^{-1}$
}
\label{VV_747_HI}
\end{figure}

For the total integration time of 234 minutes (ON+OFF source)
the r.m.s. noise is 4~mJy. The galaxy is detected with S/N ratio of 12.
The spectrum is presented in Fig.~\ref{VV_747_HI}.
The line shape is close to gaussian and typical of low-mass galaxies.
Its width at 50 \% of peak W$_{0.5}$ = 90 km~s$^{-1}$
is close to the median value of the distribution derived for the statistical
sample of BCGs from the zone of the Second Byurakan Survey (SBS)
by Thuan et al. (\cite{Thuan99}).
The line width at 20 \% of peak  W$_{0.2} = 117$ km~s$^{-1}$ is
very close to that  of VV~432. This is important in the following discussion
about the applicability
of the Tully-Fisher relation (\cite{Tully77}, hereafter TF)  for the
distance problem of the latter galaxy.

The H{\sc i}-mass of VV~747 of 0.8$\times$10$^{8}$ M$_{\odot}$,
is typical of low luminosity gas-rich galaxies, like the ratio of the 
hydrogen mass over the blue luminosity: M(H{\sc i})/L$_B = 1.0$.

\section{Properties of individual galaxies}
\label{Individual_prop}

\subsection{VV~432 = IC~3105 = UGC~7326}

The first long-slit spectrum of this chain-like galaxy has been obtained along
the major axis by Arkhipova et al. (\cite{Arkhipova81}).
They revealed high-excitation spectrum of
H{\sc ii}-region in the  NE region. The amplitude of the rotation curve 
has been  found to be relatively
small (less than about 60 km~s$^{-1}$).


Our long-slit spectrophotometry allowed to detect  the NE
H{\sc ii}-region (knot ``a'' on Fig.~\ref{VV432_fig})
[O{\sc iii}]-line $\lambda$\,4363~\AA\, with a
good S/N ratio in both spectra. The first spectrum, obtained on February 12,
1999 covered the range 3600--8000~\AA\, and the spectra taken on April 24,
1999 with a dispersion 1.2~\AA/pixel, covered two separate ranges
4000--5200~\AA\, and 6000--7200~\AA.  Table~\ref{Tab3} presents 
the data for all measured emission lines.
According to this table, the relative intensities
of strong oxygen and hydrogen lines in both  spectra
are the same within 3\%, in accordance with their internal errors.
For the abundance derivation we supposed that, for the April spectra, the
relative intensity of the [O{\sc ii}]-line $\lambda$\,3727~\AA\, and of the
H$\beta$ line to be equal to those derived from the February spectrum.

Oxygen abundance determination gives consistent values for both data sets,
so we accept the weighted mean of the two independent measurements. The
resulting value of 12+log(O/H) =
7.58$\pm$0.06 obtained for this H{\sc ii}-region appear to be very low.

Since the [O{\sc iii}]-line $\lambda$\,4363~\AA\, was not detected in other
emission-line knots, the oxygen abundances for knots ``b'', ``c'' and ``d'' were
estimated by the same empirical method mentioned in
Sect.~2.3.
The most uncertain parameter -- the mean gas density -- was assumed to
be close to that found for knot ``a'', what is about 1 atom~cm$^{-3}$
(the value, derived
from the consistency of oxygen abundances, determined by classical and
empirical methods). Oxygen abundances, found by the empirical method
for this value of N$_e$ for three knots, are consistent with the estimates,
obtained for knot ``a'', taking into account the uncertainties
of parameter R$_{23}$ and possible variations of the mean density
N$_e$ within the factor of 2.


This galaxy is
elongated (the axial ratio of $b/a$~$\approx$~1/3
according to LEDA)\footnote{LEDA is
the acronym of Lyon-Meudon Extragalactic Database, http://leda.univ-lyon1.fr}, 
which is not untypical
for dwarf galaxies. It looks like an edge-on disk, bent on
both NE and SW edges, which probably indicates a tidal action from some
other galaxy. The bright H{\sc ii}-region on the NE edge appears to be
outside of the main body of the galaxy, and its connection to the disk is not
evident, although a smooth velocity distribution along the galaxy
favours its interpretation as a single object.
Another possibility which cannot be excluded is that the outer knots
are large SF regions at the ends of spiral arms, seen not exactly edge-on.
However due to the rather small luminosity of VV~432, the presence of
spiral arms is quite improbable.

Our long-slit spectrum with the dispersion of 1.2~\AA/pixel allows to
derive the
rotation curve along the ``disk'', using the  H$\alpha$ line data
(see Fig.~\ref{VV432_fig}b).
A velocity gradient  is well seen over the whole disk; there is no sign
of flattening of the velocity curve on both NE and SW edges.
The mean heliocentric velocity of the galaxy is $-160$ km~s$^{-1}$.
The amplitude of the rotation curve within the whole extent of
H$\alpha$-emission of about 70$^{\prime\prime}$ corresponds to { the}
maximal rotational velocity of 40$\pm$5 km~s$^{-1}$. The expected inclination
correction is very small, since the ~disk is seen at the inclination
angle $i > 70^{\circ}$.

The uncertainties of the data points of the  Fig.~\ref{VV432_fig}b's 
velocity curve 
are small and remain in the range 4 to 15 km~s$^{-1}$.
The irregularities  which are seen on the rotation curve can then be
considered as real
ones, probably connected with the regions of  active SF.


The galaxy is situated in the direction of Virgo cluster (VC), and is also
catalogued as VCC~0241. Its negative radial velocity does not contradict its
membership to the cluster (Binggeli et al.~\cite{BPT93}).
In this case, its distance is of $\approx$~20.7~Mpc (distance modulus = 31.6~mag,
Federspiel et al. (\cite{Federspiel98})). Its maximal diameter on the
isophote $\mu_{B}$= 25$^{m}/\Box{\arcsec}$ is 111${\arcsec}$, according to
LEDA, which corresponds to  a linear size of
about 11~kpc. The extent of the H$\alpha$ emission on our long-slit spectrum
is of
70${\arcsec}$, which gives the total extent of H{\sc ii}-emission to be
$\approx$~6.8~kpc.

If VV~432 is a member of VC it appears to be the most metal-deficient known
galaxy of this cluster (see for comparison the metallicity data of the VC
BCGs in  Izotov \& Guseva (\cite{Izotov89})).
Such low metallicity implies that the galaxy presumably experienced
only one or two major SF episodes during its life. Since galaxy interactions
in clusters are very important triggers, it is difficult to understand how
VV~432 could sustain its very low metallicity in the dense VC environment.
The existence of such metal-poor gas-rich galaxy poses serious
questions on its evolution history, and
thus gives additional arguments to make independent check of its distance,
using, in particular, color-magnitude diagrams for resolved stars.

To estimate physical parameters of VV~432 we accept further
that this galaxy is in VC, at the distance 20.7~Mpc.


Since as we already noticed VV~432 resembles by its morphology an
edge-on bent disk, it is natural to check probable companion galaxies, acting
as strong enough disturbing bodies.

According to NED
\footnote{The NASA/IPAC Extragalactic Database (NED) is operated
by the Jet Propulsion Laboratory, California Institute of Technology,
under contract with the National Aeronautics and Space Administration.},
the nearest galaxies in both the projected distance
and the relative velocity are VCC~0200 (MCG~+02--31--076)  at the
angular distance 41.4$^{\prime}$ and V$_{hel}$ = 65 km~s$^{-1}$, with
B-magnitude = $14\fm69$, and NGC~4216 (UGC~7284) at the angular distance
51.8$^{\prime}$ and V$_{hel}$ = 131 km~s$^{-1}$, with B-magnitude =
$10\fm99$. If they are members of VC, their respective projected distances
relative to VV~432 are 240 and 312~kpc. NGC~4216 according
to LEDA is a giant spiral with M$_{B}$= $-21\fm6$.
If it is indeed on the same radial distance as VV~432, it really can
exert strong enough tidal action on VV~432 to trigger gravitational
instability and subsequent SF bursts. In particular, tidal generation
of shocks in gaseous disk according to Icke (\cite{Icke85}) mechanism can be
responsible for observed SF bursts (see e.g. Pustilnik et al. (\cite{Pustilnik00})
for the estimates of tidal effect  in similar situation).


\begin{figure*}[hbtp]
\vspace*{-1.0cm}
\psfig{figure=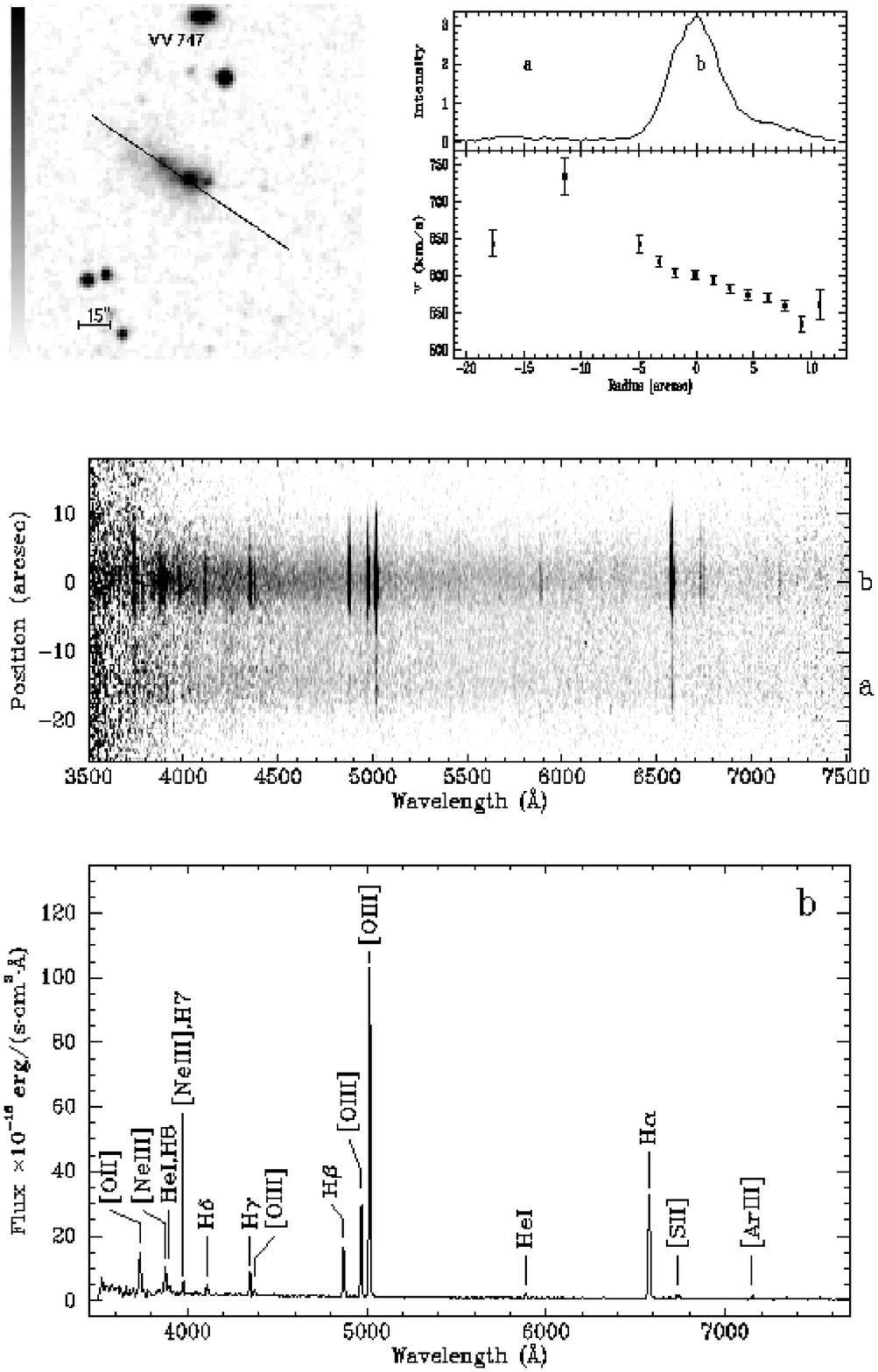,angle=0,width=18cm}
\caption[c]{
{\bf a)} DSS-1 image of galaxy VV~747 with the position of
long slit superimposed;
{\bf b)} brightness profile of H$\alpha$
along the slit and the velocity curve of VV~747;
{\bf c)} 2-D spectrum of VV~747;
{\bf d)} 1-D spectrum of the brightest SW component
}
\label{VV747_fig}
\end{figure*}


\begin{table*}[hbtp]
    \centering{
\caption{\label{Tab6} Abundances in VV~432, VV~543W and VV~747}
\begin{tabular}{lcccc} \hline
\rule{0pt}{10pt}
Value                     & VV~432 (a) (4.6 \AA/$pixel)$ &VV~432 (a) (1.2 \AA/$pixel)$ & VV~747 (a) (4.6 \AA/$pixel)$ &VV~543W (4.6 \AA/$pixel)$ \\ \cline{1-5}

T$_e$(O{\sc iii})(K)\     & 20680$\pm$1720       & 18480$\pm$3320   & 15380$\pm$840        & 7500$\pm$1500         \\
N$_e$(S{\sc ii})(cm$^{-3}$)\ &   $<$ 10          &    $<$ 10        & 110$\pm$239          &  $<$ 10               \\
12+log(O/H)\              &  7.57$\pm$0.06       & 7.62$\pm$0.13    &  7.85$\pm$0.05       &  8.5$\pm$0.1$^{(1)}$  \\
log(N/O)\                 & --1.41$\pm$0.17      & --1.50$\pm$0.34  & --1.47$\pm$0.16      & ---                   \\
log(Ne/O)\                & --0.63$\pm$0.12      & ---              & --0.78$\pm$0.10      & ---                   \\
\hline
\multicolumn{5}{l}{\rule{0pt}{10pt}(1) --- without [O{\sc iii}]-line $\lambda$\,4363~\AA\
obtained by the method described in section~\ref{Abund_determ} } \\
\end{tabular}
}
\end{table*}

\subsection{VV~543 = NGC~5275}

This galaxy was considered as a candidate dwarf galaxy (M$_{B}$= $-15\fm5$)
due to its catalog radial velocity
V = 1\,395 km~s$^{-1}$ (RC3, de Vaucouleurs et al. (\cite{deVaucouleurs91}),
cited also in both LEDA and NED, and originally obtained by Arkhipova \&
Esipov (\cite{Arkhipova79})). However colour indices of this object are
rather typical of giant E galaxies than for dIrr systems (Zasov \&
Arkhipova 2000). According to our observations, it is evident
that there was some misprint in the original work, caused the catalog velocity
of this galaxy to be in error by the factor of ten. Its real velocity,
measured from our spectrum of emission-line region on the western
periphery (hereafter VV~543W) is 14\,100 km~s$^{-1}$.

In fact, the spectrum of the central bright region of this galaxy shows, that
we have in this case an optical pair in projection. Indeed the central
part shows only the absorption lines, typical of elliptical galaxy
(e.g. Pickles (\cite{Pickles88})),
but its radial velocity is  1620$\pm$120 km~s$^{-1}$  lower than that
for the emission-line galaxy. Its luminosity M$_{B}$ = $-20\fm9$
evidences that this galaxy falls to the class of normal ellipticals. The apparent
magnitude of VV~543W is $\approx$~2\fm5
fainter (B~$\approx 17\fm7$) than that of VV~543E (this estimate
follows from the comparison of the flux near $\lambda$\,4400~\AA\, in both
galaxies), what leads to M$_{B} \sim$ $-18\fm7$. Its linear size along the
major axis $\sim$~13~kpc is quite
modest, and, owing to the typical spectrum of H{\sc ii}-region, this
western component can be considered as a bright blue compact/H{\sc ii}
galaxy. The compact object lying at about 20${\arcsec}$ to NE
from the absorption-line galaxy is a foreground star.

The extent of H$\alpha$ emission along the minor axis is traced within
8${\arcsec}$. Even after the binning in 4 pixels (which corresponds to
the seeing of
2${\arcsec}$) the velocity curve does not indicate clear gradient.
The full range of the radial velocity
is about 80~km~s$^{-1}$ with the mean value of 14\,100$\pm$30~km~s$^{-1}$.

According to NED, VV~543W has a probable companion galaxy NGP9
F324--0303806 at 5.8$^{\prime}$ with the radial velocity 13\,699$\pm$126 km~s$^{-1}$
and B = $16\fm68$. Corresponding projection distance $\approx$~300~kpc
and velocity difference 401$\pm$130 km~s$^{-1}$ are in the range
typical of wide pairs of binary galaxies (see e.g. the study
by Chengalur et al. (\cite{Chengalur93}) and Nordgren et al.
(\cite{Nordgren98})). Some weak tidal action from this galaxy can
be responsible for the enhanced  SF in VV~543W
(see e.g. Reshetnikov \& Combes (\cite{Reshet97}), and Rudnick \& Rix
(\cite{Rudnick98})).


\subsection{VV~747 = CG~798}
\label{747_mult}

This dwarf galaxy consists of two clearly separated regions embedded into the
common envelope of low surface brightness. The brighter SW
component has full size of about 15${\arcsec}$ and the fainter NE can be
traced down to about 8${\arcsec}$ ($\approx$~0.8 and 0.4~kpc
respectively). Both regions show emission-line spectrum (see
Fig.~\ref{VV747_fig}b).
The high excitation spectrum with the observed [O{\sc iii}]
line $\lambda$\,4363~\AA\, of the SW component  allows to determine oxygen
abundance using the direct measurement of $T_e$.
Our O abundance is 12+log(O/H) = 7.85$\pm$0.05  compared to the abundance 7.97
derived by Izotov \& Thuan (\cite{Izotov99}). This low metallicity is not
untypical for BCGs.

Smooth velocity curve along the slit shows a small but clear slope across the
SW component in the region of high S/N ratio of H$\alpha$ with the total
extent of about 15${\arcsec}$ and the velocity range from 540 to
630 km~s$^{-1}$, consistent with the maximal rotation velocity of about
45 km~s$^{-1}$ and mean radial velocity of 585 km~s$^{-1}$. The galaxy body
is well elongated, so the inclination correction does not seem to be larger
than 20$\div$30\%. In the NE component the velocity can be measured
only in two independent points, showing significant scattering,
consistent with their internal uncertainties. The mean velocity of NE
component, taking from these two points, is about 690$\pm$42 km~s$^{-1}$.

Huchra et al. (\cite{Huchra95}) presented  the values of the observed
velocities
for two components of this galaxy, designated as SW and NE, but
their  coordinates given in the paper, are  the same for both of them.
If we accept that
they observed the same components, our results are consistent with theirs
(V(SW) = 621$\pm$32 and V(NE) = 665$\pm$44 km~s$^{-1}$) within the cited
uncertainties.

Since the two separate velocity points for the NE component do not show any
clear gradient, which would indicate its independent rotation, and  the
mean velocity of the NE component matches well the continuation of the
velocity curve for the SW component, there is no reason to consider this
system as two different galaxies in collision. The current data favour the
interpretation
of this system as a single galaxy with two super-giant H{\sc ii}-regions in
different excitation stages. In this case the systemic velocity of this
single galaxy is about 620 km~s$^{-1}$ with the full velocity range of 540
to 710 km~s$^{-1}$. Both these values are quite well consistent with
the parameters of H{\sc i}-profile of VV~747 described and shown in
section~\ref{HI_res}.  The latter full velocity range corresponds to
the maximal rotational velocity of 85 km~s$^{-1}$.

The better knowledge of the velocity curve in the region of the NE
component is necessary to exclude completely the hypothesis of merger of two
dwarf galaxies. At least one indirect argument favours this interpretation.
If TF relation holds for this galaxy, its H{\sc i} mass and blue luminosity
are more consistent with the narrower width (W$_{0.2}$ = 70--80 km~s$^{-1}$),
than with the observed one in the integrated H{\sc i}-profile. It can
indicate that
we observe the sum of two ``narrow'' H{\sc i}-profiles displaced at about
40--50 km~s$^{-1}$ each from other.
Only 21~cm line mapping of H{\sc i} velocity field and/or very high quality
H$\alpha$ velocity data can help to make a final interpretation
of this system.

The search in the NED resulted in the bright companion galaxy NGC~3432 (= Arp~206
= VV~011)  at the angular distance 67$^{\prime}$ with B = $11\fm67$ and
the radial velocity 616 km~s$^{-1}$, very close to that of VV~747.
This galaxy, classified as SB(s)m (LINER H{\sc ii}), has M$_{B}$ = $-18\fm3$.
At the projected distance $\approx$~190~kpc the tidal
action of this galaxy to VV~747 may also be strong enough to trigger
SF in this tiny dwarf.

\section{Discussion and conclusions}
\label{Conclusions}

One of the important questions concerning the nature of the ``nest'' and
``chains'' of VV-galaxies
is their evolution status. Current study as well as several previous
publications demonstrate that many of the low-luminosity VV-galaxies
are relatively nearby irregular galaxies with several bright knots of
enhanced SF. Their position in two-colour diagram
indicates a presence of SF burst in many of them (Zasov
\& Arkhipova \cite{Zasov00}).
They roughly follow so called luminosity -- metallicity relation:
the least luminous galaxies show in general smaller values of $O/H$.

VV~432 has a very low heavy-element abundance. Its $O/H$ is among the
lowest ten values of the most metal-deficient BCGs out of more than one thousand
BCGs/H{\sc ii}-galaxies known up-to-now. It may be considered
as an example of non-evolved galaxy.
Analysis of the empirical correlations suggests that dwarf galaxies
with 12+log(O/H) $<$ 7.6 can currently experience only the first in their
history episode of SF (Izotov \& Thuan \cite{Izotov99}).
Therefore VV~432 with 12+log(O/H) = 7.58 is very good
candidate for more detailed study. If it is situated in Virgo cluster, it
will be the most metal deficient galaxy of this aggregate, being even less
chemically evolved than another well-known metal-poor H{\sc ii}-galaxy
in the direction of Virgo cluster H{\sc i}~1225+01 with 12+log(O/H) = 7.66
(Salzer et al. \cite{Salzer91}; Chengalur et al. \cite{Chengalur95}).
One of the possible ways to resolve  the
dilemma of radial distance to VV~432 is a detection of brightest stars
and construction of their color-magnitude diagram.

The galaxies we discuss have rather close neighbours.
Icke (\cite{Icke85}) first has drawn attention to the importance of
relatively weak interactions to trigger gravitational instability in gas disks
via generation of shocks. Many observational evidences for the important role
of weak interactions to trigger SF were obtained since that time, including
the detection of low mass H{\sc i}-companions of nearby H{\sc ii}-galaxies
(Chengalur et al. {\cite{Chengalur95}; Taylor et al. \cite{Taylor93},
\cite{Taylor95}; Taylor \cite{Taylor97}) and optical faint companions of BCGs
(Pustilnik et al. {\cite{Pustilnik97}). Recent results on late spirals by
Reshetnikov \& Combes (\cite{Reshet97}) and Rudnick \& Rix (\cite{Rudnick98})
also suggest the importance of weak interactions to modulate SF history
in these galaxies.

From the observational data discussed above some preliminary conclusions
can be drawn:

\begin{itemize}
\item

Spectrophotometry of VV-galaxies shows that low luminosity representatives
of this sample are in general metal-deficient objects, and in this aspect
they are similar to dwarf irregular galaxies.

\item

The extremely metal-deficient galaxy VV~432 (12+log(O/H) = 7.58) is probably
the least evolved known member of Virgo cluster.

\item

The system VV~543 radial velocity cited in the RC3 catalog and other
databases is wrong. This object consists of two galaxies with discordant
redshifts
(a unique example among the galaxies of this type!), and probably presents
an optical pair. VV~543W is an H{\sc ii}-galaxy with
the radial velocity 1620 km~s$^{-1}$ higher than that of absorption-line
E-type galaxy VV~543E.

\item

VV~747 is probably a single dwarf galaxy rich of neutral hydrogen.

\item

The presence of massive companion galaxies at the distances of few hundred
kpc from the studied VV-objects with enhanced  SF rate is probably
indicative of the important role of weak interactions to trigger SF activity
at least in some fraction of low mass VV galaxies.

\end{itemize}

\begin{acknowledgements} We thank D.Makarov (SAO) for providing us with his
MIDAS package to derive velocity curve on long-slit spectra.
SAO authors appreciate the partial financial
support from the RFBR grant No. 96-02-16398
One of the authors (A.Z.) thanks RFBR
(grant 98-02-17102) and Federal program ``Astronomy'' for financial support.
We have made use of the NASA/IPAC Extragalactic Database (NED),
which is operated by the Jet Propulsion Laboratory, California Institute
of Technology, under contract with the National Aeronautics and Space
Administration, and of the 
Lyon-Meudon Extragalactic Database, http://leda.univ-lyon1.fr.

\end{acknowledgements}


\end{document}